\documentclass[letterpaper, 10 pt, conference]{ieeeconf}  

\IEEEoverridecommandlockouts                              

\usepackage{graphics} 

\usepackage{graphicx}

\title{\LARGE \bf
Social-Qwen: From Individual Nonverbal Cues and Emotion to Multiparty Social Dynamics Understanding with Instruction Tuning  
}

\author{Tung The Nguyen$^{1}$ and Jouh Yeong Chew$^{1}$
\thanks{$^{1}$The authors are with Honda Research Institute Japan Co., Ltd.
        {\tt\small tung.nguyen@jp.honda-ri.com}, {\tt\small jouhyeong.chew@jp.honda-ri.com}}
}

\begin{document}

\newcommand{\methodname}{Social-Qwen\space}

\maketitle
\thispagestyle{empty}
\pagestyle{empty}

\begin{abstract}

Effective participation in multiparty scenarios requires robots to move beyond individual toward understanding group-level social dynamics, which are inherently complex due to the interplay of nonverbal cues, internal states, and interaction context. Existing approaches often rely on end-to-end deterministic models, while recent state-of-the-art methods such as large Vision-Language Models (VLMs) address this issue to some extent but remain limited by their size and computational cost for real-time applications. Moreover, both approaches are constrained by the scarcity of multiparty interaction data and annotations, which describe how individual nonverbal cues and emotional states contribute to social dynamics which describe collective outcomes such as group engagement. We hypothesize that explicitly modeling individual-level states is essential for accurate group-level understanding. To this end, we present Social-Qwen, a two-stage framework that first analyzes each participant’s nonverbal cues and emotions, then infers group-level engagement using instruction-tuned representations. To mitigate the lack of individual annotations in group datasets, we employ knowledge distillation to transfer supervision signals. Experiments on the OUC-CGE dataset show that Social-Qwen significantly outperforms prior end-to-end baselines and achieves state-of-the-art performance in group engagement analysis, demonstrating the promise of instruction tuning for scalable social intelligence in robots. We further evaluate robustness by testing generalization to (1) an in-house dataset spanning multiple social activities and (2) estimating other social dynamics such as group harmony. Results suggest consistent performance, highlighting Social-Qwen as a promising approach toward real-time social intelligence for intelligent agents. 

\end{abstract}

\section{INTRODUCTION}
\label{sec:introduction}

With the rapid development in technologies and engineering, the application of robots for facilitation of multiparty interactions \cite{multiparty-facilitation, nguyen2023study, tang2024assisting} starts to receive more attention from the research community. In multiparty activities, it is critical for robots to understand social interaction dynamics \cite{tgn-pl,social-dynamics-tgn} encompassing additional emergent properties such as cooperation, role distribution, and harmony. Group engagement can be viewed as a key dimension of social interaction dynamics, capturing the collective attentional and participatory state of a group. Specifically, group engagement refers to the level of focus and attention of the whole group of participants towards the activity that is being conducted. Therefore, the ability to measure the group engagement level is crucial to facilitate the group effectively. With the group engagement analyses, the robot can achieve better understanding of the multiparty situation and group dynamics, then reacts appropriately to manage the group interactions.

The problem of automatically estimating individual's level of engagement when performing a given task has been investigated extensively. The majority of existing studies, particularly those in recent literature, have leveraged machine learning techniques for engagement analysis \cite{tang2025facial, buono2023assessing, garcia2024study}. In addition, those studies usually utilizes visual cues and linguistic features such as spoken utterances to predict the person's level of focus on their given tasks. Finally, existing dataset for engagement analysis primarily focus on activities in classroom or working office, due to the high practicalities and usefulness of engagement in these scenarios \cite{garcia2024study, hsia2022student, tang2025facial, speer2023turning}. Unfortunately, despite these advances in individual engagement understanding, studies in group engagement remain at early stages. To our knowledge, Lu et al. (2025) is the first to define group engagement and create a dataset for analyzing group engagement using visual information in a classroom setting \cite{lu2025video}. In this study, the authors utilize convolutional neural network (CNN) as the model to perform prediction of group engagement.  

\begin{figure}[t]
    \centering
    \includegraphics[width=1.\linewidth]{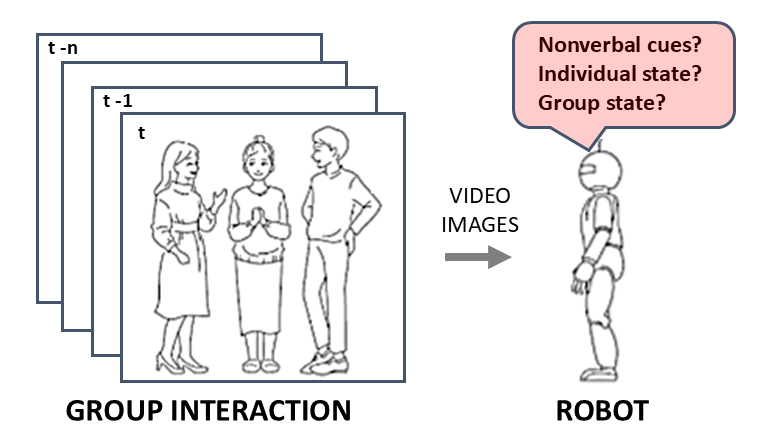}
    \vspace{-10pt}
    \caption{Overview of the problem formulation. Given a sequence of video images of a multiparty interaction, intelligent agents must infer the corresponding social dynamics such as engagement based on the understanding of individual nonverbal cues and emotion.}
    \label{fig:overview}
\end{figure}

While previous study demonstrated that end-to-end method achieved very good performance in group engagement analysis, we hypothesize that individual behaviors and other non-verbal cues are strongly correlated to group engagement level. This is inline with the preceding study \cite{internal-state-jcogsci} which suggested the individuals' internal cognitive states during group interactions are associated with nonverbal cues. For example, in the studying scenario, if all the students shows clear signs of boredom and they are not paying attention towards the teacher and learning materials, then we can conclude that the group engagement is low. Therefore, the accuracy of predicting group engagement level can be further improved by analyzing individual cues, then utilize the analysis results to predict the group engagement level. This two-step individual-to-multiparty approach allows thorough analysis of the group situations and accurately estimate the group engagement level. An illustration of this process is shown in Figure \ref{fig:overview}. However, within existing group engagement dataset, ground-truth labels for these individual analysis are not available. To overcome this issue, we leverage knowledge distillation approach and use a large visual language model (VLM) to generate pseudo-labels of the individual cues. Subsequently, these pseudo-labels are utilized to finetune a smaller VLM for individual-level analysis.

Our contributions are as follows:
\begin{itemize}
    \item We propose \methodname, a novel two-step approach to tackle the problem of group engagement analysis. The framework first analyzes individual cues from each participant in the group. Then, utilizes the analyses results to effectively predict the group engagement level, demonstrating transition from individual towards group-level understanding of social interaction dynamics.
    \item We conduct an extensive experiments and show that the proposed \methodname significantly outperforms several strong baselines, achieving state-of-the-art performance on a publicly available dataset. 
    \item We evaluate the robustness of \methodname using an in-house dataset spanning across multiple social activities to test a different task. The results suggest that the model exhibits strong generalization and zero-shot learning ability, enabling it to infer other group dynamics that it was not explicitly trained on. The in-house dataset will be made public.   
    
\end{itemize}

\section{RELATED WORK}
\label{sec:related_work}
\begin{figure*}[!t]
    \centering
    \includegraphics[width=1.05\linewidth]{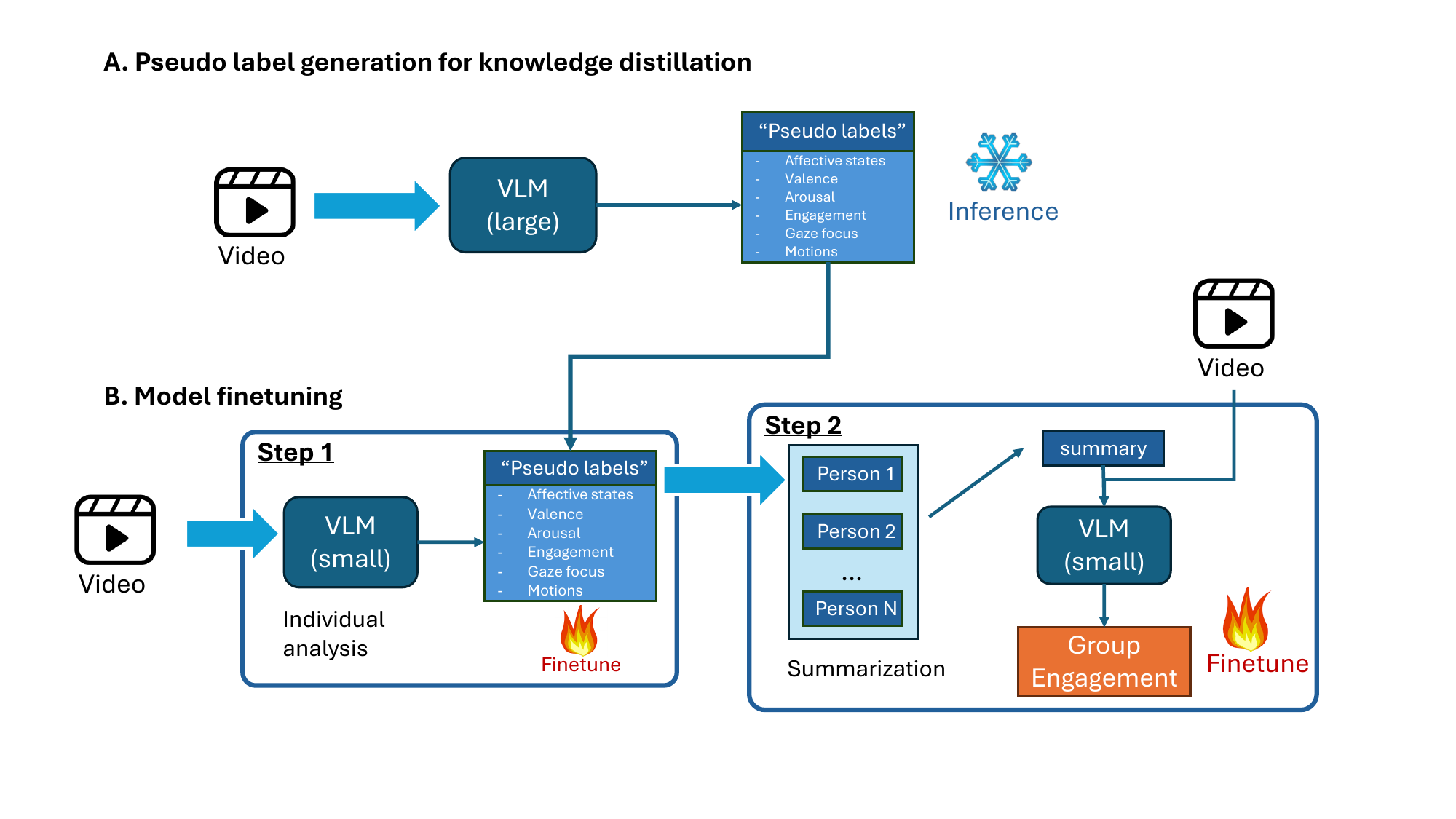}
    \vspace{-45pt}
    \caption{Overview of the proposed approach. In this paper, "VLM (large)" refers to models on the scale of dozens of billions of parameters or more, such as Qwen2.5-VL-72B-Instruct. Meanwhile, "VLM (small)" refers to more compact models such as Qwen2.5-VL-3B-Instruct.}
    \label{fig:model_arch}
\end{figure*}

\textbf{Engagement analysis}. 
Recent research has made significant strides in analyzing engagement through facial expressions, with most of the studies focused on student engagement in learning or work environments. These studies leverage machine learning and affective computing to provide objective, real-time insights that were previously reliant on subjective self-reporting.

One study proposed a model using Long Short-Term Memory (LSTM) networks to predict engagement levels from facial action units, gaze, and head poses \cite{buono2023assessing}. By comparing the model's predictions with students' self-reported engagement after an online lecture, the study found a correlation, particularly with the emotional dimension of engagement. The analysis revealed that facial movements and head poses are positively correlated with engagement, while gaze showed an inverse correlation. These findings suggest that automated analysis of facial behavior can provide valuable, objective insights into student engagement during remote learning.

Another recent study proposes a comprehensive system for analyzing student engagement in online courses by combining the detection of both macro-expressions (obvious, longer-lasting emotions) and micro-expressions (subtle, fleeting facial movements). The researchers developed a system that uses a deep learning model to evaluate both emotional and behavioral dimensions of engagement from video streams. Their findings indicate that the emotional dimension, captured through detailed facial expression analysis, has a greater impact on student engagement than behavioral aspects like note-taking. This approach provides teachers with more objective and nuanced metrics to assess student participation and adjust their teaching strategies in an online setting.

Traditionally, engagement analysis in office environment is based on long-term information such as social media posts, online chat logs, or reports collected through survey \cite{athukorala2020business, tanaka2021estimating, shami2015inferring}. Real-time analysis of engagement level in office using nonverbal cues such as facial expressions have started to gain attention recently. Chang et al. (2018) proposed a novel multimodal engagement estimation model based on ensembling of clustering methods and attention-based RNN \cite{chang2018ensemble}. The results from this study indicate that body language alongside facial expressions, significantly improves the accuracy of engagement prediction. This demonstrates the value of using a holistic set of visual cues rather than relying on facial analysis alone. A more recent research from Zhu et al. (2020) proposed a multi-rate attention mechanism and achieved more accurate prediction of the employee's engagement state. 

\textbf{Group engagement}.
Lu et al. (2025) proposed the first study on group engagement in a classroom scenario \cite{lu2025video}. Based on the ICAP framework \cite{chi2014icap}, the author collected and constructed a dataset for group engagement analysis task. In particular, the dataset contains recording of several groups of students with various sizes, each consists of six to eight students, participated in a lecture with predefined structure. Each lecture contains introduction (10 minutes), knowledge lecture (20 minutes), group task (25 minutes) and debriefing (15 minutes). The recordings are segmented into 10-second intervals as data samples. Each segment is annotated with one of three classes, low, medium, or high by an expert. Additionally, self-reported engagement levels are collected from each students. Finally, the class labels are discussed and finalized considering both self-reported results and annotations. The authors implemented several baselines with CNN end-to-end models and achieved more than 90$\%$ accuracy. This result demonstrated the practicality of applying machine learning methods for automatically estimating group engagement.   

\textbf{Knowledge distillation}.
Knowledge distillation is a deep learning technique proposed by Hinton et al. (2015) \cite{hinton2015distilling}. In principle, this approach uses the "teacher-student" paradigm where a smaller "student" model is trained to mimic the output logits (soft labels) of a larger, more complex "teacher" model or an ensemble of models. Knowledge distillation was successfully applied to large language models (LLM), allowing smaller model to achieve performance close to that of much larger ones \cite{sanh2019distilbert, li2023prompt, li2025bayeskd}. More recently, the application of knowledge distillation in VLMs has also been investigated and showed very promising results \cite{fang2021compressing, jang2025vl2lite, cao2025move}. In this study, we utilize knowledge distillation approach to overcome the issue of unavailability of individual labels in the group engagement dataset in \cite{lu2025video}. 




\section{METHOD}
\label{sec:method}
\subsection{Overview}
\label{ssec:overview}
Instead of using the classification approach in \cite{lu2025video}, we formulate the problem of group engagement analysis as a text generation problem and utilized VLM as the base model. In particular, let us denote video features that were extracted from the video as $V$ and linguistic features from instruction prompt text as $T$. The output of \methodname is given by:
\begin{equation}
    result = f_{VLM}(V,T)
\end{equation}
The generated result text is in dictionary format, for example \textit{\{``group engagement": ``high"\}}. This format allows easy extraction of the engagement value from the output text. Although the generative approach is characterized by greater complexity and typically exhibits lower performance than classification modeling, it offers much better robustness and generalization. 

The overview of our proposed approach is illustrated in Figure \ref{fig:model_arch}. We separate the prediction of group engagement into two steps: individual analysis and summarization. The premise of our work is that, akin to human intuition, the analysis of group dynamics such as group engagement should begin at the individual level. Therefore, \methodname is designed to first assess the distinct behavioral expressions of each person (individual analysis) then proceed with summarization of the individual cues to estimate the group engagement. The following subsections detail the model finetuning procedure for our proposed approach.

\subsection{Individual Analysis}
\label{ssec:individual_analysis}
We utilize a large VLM model as the ``teacher" as generate individual pseudo labels, which include affective states, valence, arousal, engagement, gaze (visual) focus, and motions of each participant in the multiparty activity. These pseudo labels are then utilized to finetune smaller ``student" model to replicate the individual analysis performance of the ``teacher" model. 

\begin{itemize}
    \item \textit{Affective states}. Affective states represent temporary, subjective experiences of feeling or emotion, encompassing moods, stress levels, and specific emotional responses of each individual. Therefore, they are important information for group engagement analysis. We consider 14 affective states: Anger, Anxiety, Boredom, Confusion, Contempt, Curiosity, Disgust, Eureka, Fear, Frustration, Happiness, Neutral, Sadness, Surprise. These classification labels are based on a study by D'Mello et al. (2010) \cite{d2010monitoring}. 
    \item \textit{Valence and Arousal}. In addition to the classes above, we also consider valence and arousal dimensions of affective states. In this study, the values of valence and arousal are within the range $[-4,4]$.
    \item \textit{Visual focus}. Visual focus, the direction of a participant's gaze, is another crucial individual cue for assessing group engagement. For example, if most of the participants are not paying attention to the studying materials or the tutor, it's a clear sign that the group has low engagement. We specify the visual focus results with the following choices: "looking at the teacher", "looking at the board", "looking at their study materials", "looking at personal items", "looking at other students", or "not looking at anything". 
    \item \textit{Motions}. Motion cues, which encompass the gestures and bodily movements of a participant in a group activity, are highly generalizable and useful not only for analyzing group engagement but also for other group dynamics understanding tasks.
    \item \textit{Individual engagement}. Individual engagement, which indicates the level of focus of each participant toward the group activity, is directly correlated to the group engagement level. Thus, we also include this cue in the individual analysis. 
\end{itemize}

\subsection{Summarization}
\label{ssec:summarization}
For the summarization step, we finetune a model to predict group engagement based on the input video and the aggregated summary of individual analyses. In the training phase, the analyses summary is created directly from the pseudo labels. During inference, however, this summary is constructed from the individual cues produced by the "student" model for each participant. Note that we can finetune the same model for both individual analysis and summarization, hence reduces the storage requirement for the VLMs. There are various methods to summarize the individual analyses, such as using template-based method or text summarization model. In our study, we use simple concatenation of the generated texts from individual analysis for faster processing. 

Alternatively, the individual analysis and summarization steps could be modeled as a single, unified phase, similar to that of chain-of-thought studies. With this approach, the individual analysis can be viewed as the reasoning process before giving the result of group engagement analysis. A key disadvantage of this approach is that the target text is much longer, making it more difficult and potentially takes significant time to train the model. In addition to that, despite this approach allows performing inference in one pass, it actually takes more time compared to our two-step approach. With \methodname, we can apply mini-batch inference or other parallel processing methods, allowing simultaneous generation of individual analysis for each participant. In other hand, chain-of-thought approach needs to generate the analysis for each participant sequentially, thus, significantly more time-consuming. Therefore, our proposed approach allows much faster inference time and is more suitable for practical applications and deployment to robots. 

\section{EVALUATION}
\label{sec:evaluation}
We conducted an experiment to verify the following hypotheses.
\begin{itemize}
    \item The model trained by our proposed two-step approach outperforms standard approach, achieving state-of-the-art results on the OUC-CGE dataset \cite{lu2025video}.
    \item Our proposed approach allows better generalization and robustness compared to the standard approach. The models trained on OUC-CGE significantly outperform baseline models in the task of group engagement and group harmony analysis in our in-house dataset. 
\end{itemize}
In addition, we performed an ablation study to assess the effects of each type of individual cues to the model's performance.
\subsection{Experiment Settings}
\label{ssec:experiment_settings}
For our experiments, we compared the performance of finetuned models that are based on Qwen2.5-VL-3B-Instruct, a prominent foundation Vision-Language Model (VLM) selected for its effective balance of a manageable size and robust performance capabilities \cite{bai2025qwen2}. The details of each model are as follows.
\begin{itemize}
    \item \textit{Standard approach}. This is a baseline model that utilize end-to-end approach.
    \item \textit{Proposed - all}. This is a model based on \methodname and uses all six individual cues: affective state, valence, arousal, visual focus, motions, and individual engagement.
    \item \textit{Proposed - affective}. This is a model based on \methodname and uses affective individual cues, which includes affective state, valence, and arousal. These affective cues represent the internal states of the participants. 
    \item \textit{Proposed - behavior}. This is a model based on \methodname and uses behavioral (non-verbal) individual cues, which includes visual focus and motions.
    \item \textit{Proposed - engagement}. This is a model based on \methodname and only uses individual engagement.
\end{itemize}
In our experiments, all ``Proposed" models contains two separate models for individual analysis and summarization. We used Qwen2.5-VL-72B-Instruct to generate pseudo labels for knowledge distillation phase. In addition to the finetuned models, we also include Qwen2.5-VL-3B-Instruct and Qwen2.5-VL-72B-Instruct as baseline models for comparison.
\begin{table}[hb]
    \centering
    \caption{Dataset statistics.}
    \label{tab:dataset_statistics}
    \begin{tabular}{|c|c|c||c|}
        \hline
        \multicolumn{4}{|c|}{\textbf{OUC-CGE Dataset}} \\ 
        \hline
        \textbf{Train} & \textbf{Development} & \textbf{Test} & \textbf{Total} \\
        \hline
        5962 & 751 & 749 & 7462\\
        \hline
    \end{tabular}
    
    \vspace{10pt}
    \begin{tabular}{|c|c|c||c|}
        \hline
        \multicolumn{4}{|c|}{\textbf{Our In-house Dataset}} \\ 
        \hline
        \textbf{Discussion} & \textbf{Debate} & \textbf{Social game} & \textbf{Total}\\
        \hline
        68 & 128 & 122 & 318\\
        \hline
    \end{tabular}
\end{table}

\begin{table*}[ht]
    \vspace{10pt}
    \centering
    \caption{Group engagement analysis results on OUC-CGE dataset.}
    \label{tab:ouc_engagement_results}
    \begin{tabular}{|l||c|c|c|c|c|c|c|c|}
        \hline
        \textbf{Model} & \textbf{Accuracy} & \textbf{Pre. (macro)} & \textbf{Rec. (macro)} & \textbf{F1 (macro)} & \textbf{Pre. (weighted)} & \textbf{Rec. (weighted)} & \textbf{F1 (weighted)}\\
        \hline
        Qwen2.5-VL-3B-Instruct & 0.3182 & 0.6327 & 0.3577 & 0.2576 & 0.6600 & 0.3182 & 0.2638\\
        \hline
        Qwen2.5-VL-72B-Instruct & 0.5273 & 0.5714 & 0.5682 & 0.5162 & 0.6021 & 0.5273 & 0.5031\\
        \hline
        Lu et al. (2025) \cite{lu2025video} & 0.9780 & 0.9760 & 0.9780 & 0.9770 & $\_$ & $\_$ & $\_$ \\
        \hline
        Standard approach & 0.9680 & 0.9699 & 0.9595 &0.9639 & 0.9684 & 0.9680 & 0.9676 \\        
        \hline
        Social-Qwen - all & 0.9893 & 0.9880 & 0.9877 & 0.9878 & 0.9893 & 0.9893 & 0.9893 \\
        \hline
        Social-Qwen - affective & 0.9893 & 0.9895 & 0.9873 & 0.9884 & 0.9894 & 0.9893 & 0.9893 \\
        \hline
        Social-Qwen - behavior & \textbf{0.9960} & \textbf{0.9962} & \textbf{0.9947} & \textbf{0.9954} & \textbf{0.9960} & \textbf{0.9960} & \textbf{0.9960} \\
        \hline
        Social-Qwen - engagement & 0.9933 & 0.9930 & 0.9922 & 0.9926 & 0.9933 & 0.9933 & 0.9933 \\
        \hline
    \end{tabular}
\end{table*}
\textbf{Datasets}. We evaluated and compared the models' performance on two datasets. The first is the OUC-CGE dataset, which was created by Lu et al. (2025) \cite{lu2025video} for group engagement in classroom scenario. Details of this dataset have been describe in Section \ref{sec:related_work}. The second dataset is our in-house dataset which contains video recordings and transcripts of three multiparty activities: a discussion about a movie that was shown to all participants, a debate about a given topic, and a social quiz game. Each sample is 10 seconds long and is manually annotated with group harmony and group engagement labels by three experts. We will provide comprehensive descriptions of this dataset in a future work. The statistics of these two datasets are shown in Table \ref{tab:dataset_statistics}. The top table shows number of samples for training, development, and test set of OUC-CGE dataset. The lower table show the number of samples for each activity in our in-house dataset. We use this dataset solely for zero-shot evaluation, therefore, all the numbers reported in this table are all used as test set. 

\textbf{Environments and training hyper-parameters}. 
For reproduction of the experiment, we provide the environment and model training hyper-parameter settings in Table \ref{tab:experiment_settings}. 
\begin{table}[htbp]
    \centering
    \caption{Experiment settings.}
    \label{tab:experiment_settings}
    
    \begin{tabular}{|l|l|} 
        \hline
        \multicolumn{2}{|c|}{\textbf{System Information}} \\ 
        \hline
        
        \hline
        Operating System         & Ubuntu 22.04.5 LTS \\
        Hardware        & NVIDIA A100-SXM4-80GB \\
        CUDA Version         & 12.8 \\
        \hline
        \hline
        \multicolumn{2}{|c|}{\textbf{Model Training Settings}} \\ 
        \hline
        Base model & Qwen2.5-VL-3B-Instruct \\
        Mini-batch size & 96 \\
        Number of epochs & 10 \\
        Frame per second (FPS) & 4 \\
        Learning rate & $1e^{-5}$\\
        Video resolution & 1280 x 720\\
        \hline
    \end{tabular}
\end{table}

For finetuning the models, we used the publicly available scripts at \emph{https://github.com/2U1/Qwen2-VL-Finetune}. All models were trained using the same settings in Table \ref{tab:experiment_settings}. Rather than a partial finetuning approach such as LoRA, all model parameters were updated during finetuning. We set the frame sampling rate (FPS) to 4, following the observations reported in \cite{lu2025video}. The input video resolution is 1280x720, which is the resolution of the videos in the OUC-CGE dataset. The source code, instruction prompts, and corresponding model weights in this experiment will be made publicly available upon publication.

\subsection{Group Engagement Evaluation on OUC-CGE}
\label{ssec:main_results}
Table \ref{tab:ouc_engagement_results} shows the results of group engagement analysis on OUC-CGE dataset. In this table, "Pre.", "Rec.", and "F1" denote precision, recall, and F1-score, respectively. All metrics are reported as both macro and weighted averages. With macro average, the score is the mean of the scores of each class, "low", "medium", and "high", hence, it is suitable to evaluate whether a model is good across all classes, including minority ones, with smaller support (number of instances in the dataset). On the other hand, weighted average is the mean weighted by the class's support. Thus, it is good at  
understanding the model's average performance per instance. In other literature, weighted average is also referred to as micro-average measures. Qwen2.5-VL-3B-Instruct and Qwen2.5-VL-72B-Instruct's results are zero-shot inference, while the other models are finetuned on OUC-CGE dataset. For Lu et al. (2025) model, we report the results of the best model report in the original paper, which is a model that uses convolutional neural network with a ``slow" approach. As the original study by Lu et al. \cite{lu2025video} did not report weighted average scores, the corresponding entries in Table \ref{tab:ouc_engagement_results} are left blank. 

\textbf{Main results}. The results indicate that even a large model such as Qwen2.5-VL-72B-Instruct significantly underperforms on the task of group engagement analysis. This outcome can be attributed to the nature of VLM pre-training; although these models are exposed to vast amounts of visual and text data, their objectives are primarily focused on generic visual understanding such as activity recognition or generic question answering. For tasks that require thorough analysis and understanding of multiparty scene such as group engagement, they tend to have poor performance as shown in Table \ref{tab:ouc_engagement_results}.

On the other hand, we can see that all the proposed models show significantly improvements over the baselines in all reported metrics, surpassing the results reported in \cite{lu2025video} and achieving near-perfect performance on the OUC-CGE dataset. These results support the first hypothesis that information from individual analyses helps improving performance of predicting group engagement level. However, the results also indicate that the OUC-CGE dataset presents a limited challenge to current models. This lack of difficulty can be attributed to the simplified design of the classroom interaction scenarios designed by \cite{lu2025video}.  
\begin{table*}[ht]
    \vspace{10pt}
    \centering
    \caption{Cross-domain group engagement analysis results on our dataset.}
    \label{tab:noc_engagement_results}
    \begin{tabular}{|l||c|c|c|c|c|c|c|c|}
        \hline
        \textbf{Model} & \textbf{Accuracy} & \textbf{Pre. (macro)} & \textbf{Rec. (macro)} & \textbf{F1 (macro)} & \textbf{Pre. (weighted)} & \textbf{Rec. (weighted)} & \textbf{F1 (weighted)}\\
        \hline
        \multicolumn{8}{|l|}{\textbf{Activity 1: Movie Discussion}} \\ 
        \hline
        Qwen2.5-VL-3B-Instruct & \textbf{0.8235} & 0.2745 & 0.3333 & 0.3011 & 0.6782 & \textbf{0.8235} & \textbf{0.7438} \\
        \hline
        Qwen2.5-VL-72B-Instruct & \textbf{0.8235} & 0.2745 & 0.3333 & 0.3011 & 0.6782 & \textbf{0.8235} & \textbf{0.7438} \\
        \hline
        Standard approach & 0.6176 & 0.2837 & 0.3107 & 0.2848 & 0.6539 & 0.6176 & 0.6323 \\        
        \hline
        Social-Qwen - all & \textbf{0.8235} & 0.2745 & 0.3333 & 0.3011 & 0.6782 & \textbf{0.8235} & \textbf{0.7438} \\
        \hline
        Social-Qwen - affective & \textbf{0.8235} & 0.2745 & 0.3333 & 0.3011 & 0.6782 & 0.8235 & 0.7438 \\
        \hline
        Social-Qwen - behavior & 0.8088 & 0.2736 & 0.3274 & 0.2981 & 0.6760 & 0.8088 & 0.7365 \\
        \hline
        Social-Qwen - engagement & 0.7500 & \textbf{0.3208} & \textbf{0.3643} & \textbf{0.3405} & \textbf{0.6855} & 0.7500 & 0.7161 \\
        \hline
        \multicolumn{8}{|l|}{\textbf{Activity 2: Debate}} \\ 
        \hline
        Qwen2.5-VL-3B-Instruct & 0.8047 & 0.2725 & 0.3270 & 0.2973 & 0.6706 & 0.8047 & 0.7315 \\
        \hline
        Qwen2.5-VL-72B-Instruct & \textbf{0.8203} & 0.2734 & 0.3333 & 0.3004 & 0.6729 & \textbf{0.8203} & 0.7393 \\
        \hline
        Standard approach & 0.7500 & \textbf{0.3756} & \textbf{0.3857} & \textbf{0.3806} & \textbf{0.7346} & 0.7500 & \textbf{0.7422} \\ 
        \hline
        Social-Qwen - all & 0.8125 & 0.2730 & 0.3302 & 0.2989 & 0.6718 & 0.8125 & 0.7355 \\
        \hline
        Social-Qwen - affective & 0.8047 & 0.2725 & 0.3270 & 0.2973 & 0.6706 & 0.8047 & 0.7315 \\
        \hline
        Social-Qwen - behavior & 0.8125 & 0.2730 & 0.3302 & 0.2989 & 0.6718 & 0.8125 & 0.7355 \\
        \hline
        Social-Qwen - engagement & 0.8047 & 0.2725 & 0.3270 & 0.2973 & 0.6706 & 0.8047 & 0.7315 \\
        \hline
        \multicolumn{8}{|l|}{\textbf{Activity 3: Social Game}} \\ 
        \hline
        Qwen2.5-VL-3B-Instruct & 0.4711 & 0.1583 & 0.3333 & 0.2147 & 0.2238 & 0.4711 & 0.3034 \\
        \hline
        Qwen2.5-VL-72B-Instruct & \textbf{0.6694} & \textbf{0.4978} & \textbf{0.4658} & \textbf{0.4432} & \textbf{0.7258} & \textbf{0.6694} & \textbf{0.6413} \\
        \hline
        Standard approach & 0.6364 & 0.4325 & 0.4404 & 0.4304 & 0.6287 & 0.6364 & 0.6238 \\        
        \hline
        Social-Qwen - all & 0.5455 & 0.4281 & 0.3822 & 0.3348 & 0.6240 & 0.5455 & 0.4815 \\
        \hline
        Social-Qwen - affective & 0.4959 & 0.4276 & 0.3497 & 0.2568 & 0.6241 & 0.4959 & 0.3660 \\
        \hline
        Social-Qwen - behavior & 0.5702 & 0.4663 & 0.3994 & 0.3491 & 0.6806 & 0.5702 & 0.5026 \\
        \hline
        Social-Qwen - engagement & 0.6281 & 0.4377 & 0.4354 & 0.4262 & 0.6364 & 0.6281 & 0.6173 \\
        \hline
        \multicolumn{8}{|l|}{\textbf{Overall}} \\ 
        \hline
        Qwen2.5-VL-3B-Instruct & 0.6807 & 0.2291 & 0.3308 & 0.2664 & 0.5008 & 0.6807 & 0.5699 \\
        \hline
        Qwen2.5-VL-72B-Instruct & \textbf{0.7631} & 0.3597 & 0.3841 & 0.3553 & \textbf{0.6943} & \textbf{0.7631} & \textbf{0.7027} \\
        \hline
        Standard approach & 0.6781 & \textbf{0.3778} & \textbf{0.3906} & \textbf{0.3792} & 0.6767 & 0.6781 & 0.6733 \\        
        \hline
        Social-Qwen - all & 0.7124 & 0.3328 & 0.3508 & 0.3131 & 0.6548 & 0.7124 & 0.6398 \\
        \hline
        Social-Qwen - affective & 0.6902 & 0.3324 & 0.3371 & 0.2826 & 0.6544 & 0.6902 & 0.5939 \\
        \hline
        Social-Qwen - behavior & 0.7188 & 0.3473 & 0.3561 & 0.3180 & 0.6761 & 0.7188 & 0.6464 \\
        \hline
        Social-Qwen - engagement & 0.7253 & 0.3462 & 0.3766 & 0.3560 & 0.6607 & 0.7253 & 0.6844 \\
        \hline
    \end{tabular}
\end{table*}

\textbf{Ablation study}. 
We conducted an ablation study on OUC-CGE dataset to evaluate the contribution of each type of individual cues to the overall performance of group engagement analysis. In general, from the results in Table \ref{tab:ouc_engagement_results}, having more information does not lead to better performance, as shown by the fact that ``Proposed - all" does not improve the performance compared to ``Proposed - affective", ``Proposed - behavior", and ``Proposed - engagement". This result suggests that the cues to be used for individual analysis should be carefully selected to achieve the best performance in predicting group engagement level. From the results in Table \ref{tab:ouc_engagement_results}, despite that non-verbal behavior cues give the most generic information, they are still very helpful in group engagement analysis task. This is inline with \cite{internal-state-jcogsci} which suggested the internal cognitive states during group interactions are associated with nonverbal cues. The good performance can also be explained by the fact that VLMs are trained for these type of captioning, description generation task, hence the pseudo-labels of non-verbal cues are quite accurate. The model's strong performance is expected when provided with individual engagement information, as individual engagement is a primary determinant of the overall group engagement level. Finally, the internal states, represented by affective cues, has the lowest performance compared to other type of individual cues. This can be explained by the fact that affective states are not very strong indicator of engagement level. To illustrate, the affective state of 'happiness' does not directly determine the level of engagement. An individual can be in a state of happiness while exhibiting either high engagement (e.g., active participation and enjoyment of the group study) or low engagement (e.g., pleasantness when playing with smartphones). In addition, VLMs are not very good at affective computing, thus, the generated pseudo-labels are less accurate.

\label{ssec:model_generation}

\subsection{Model Generalization Evaluation}
We performed an evaluation using our in-house dataset to evaluate the models' ability to generalize to different domain and different task. 
\begin{table*}[ht]
    \vspace{10pt}
    \centering
    \caption{Group harmony analysis results on our dataset.}
    \label{tab:noc_harmony_results}
    \begin{tabular}{|l||c|c|c|c|}
        \hline
        \textbf{Model} & \textbf{Activity 1's MAE} & \textbf{Activity 2's MAE} & \textbf{Activity 3's MAE} & \textbf{Overall MAE} \\
        \hline
        Qwen2.5-VL-3B-Instruct & 0.7594 & 0.9375 & 1.4504 & 1.0962\\
        \hline
        Qwen2.5-VL-72B-Instruct & 0.6962 & 0.8045 & \textbf{0.9619} & 0.8417\\
        \hline
        Social-Qwen - all & 0.5982 & 0.8803 & 1.2458 & 0.9602 \\
        \hline
        Social-Qwen - affective & 2.0779 & 1.9581 & 2.8630 & 2.3309 \\
        \hline
        Social-Qwen - behavior & 0.6712 & 0.8934 & 1.0243 & 0.8961 \\
        \hline
        Social-Qwen - engagement & \textbf{0.5785} & \textbf{0.7236} & 0.9889 & \textbf{0.7944} \\
        \hline
    \end{tabular}
\end{table*}
\textbf{Cross-domain}. First, we assessed the cross-domain group engagement analysis performance of the models trained on OUC-CGE dataset on our in-house dataset. Table \ref{tab:noc_engagement_results} shows the results for each activities, which include a movie discussion, a 
debate, and a social game. Additionally, this table includes overall results calculated as a weighted average of the per-activity scores, where each activity is weighted by its number of samples.

The results in Table \ref{tab:noc_engagement_results} demonstrate a significant performance degradation for all models on our dataset compared to their performance on OUC-CGE, highlighting the challenge of cross-domain generalization. This performance gap can be attributed to the domain shift between the structured classroom setting of OUC-CGE and the more varied group activity scenarios in our dataset. In the first activity, movie discussion, ``Proposed - engagement" model achieved the best scores in four metrics, and has highest performance overall. For the debate activity, the model trained with the standard approach yielded the best performance. Finally, with the social game scenario, Qwen2.5-VL-72B-Instruct outperforms all other models under consideration. Overall, the results in Table \ref{tab:noc_engagement_results} demonstrate that our proposed models exhibit robust cross-domain generalization, despite being heavily finetuned on the OUC-CGE dataset. Specifically, our models not only outperform the Qwen2.5-VL-3B-Instruct baseline significantly but also achieve performance close to that of the much larger Qwen2.5-VL-72B-Instruct, a model known for its strong generalization capabilities \cite{bai2025qwen2}.


\textbf{Cross-task}. In addition to evaluating cross-domain performance, we also investigated cross-task capabilities of the proposed models. For this evaluation, we assessed the models' performance on the task of understanding group harmony, another important group dynamic. We define group harmony as the interpersonal level of pleasantness within the group during a multiparty activity. Particularly, the greater prevalence of positive emotions, supports, positive feedback, and agreements among the participants, the higher the level of group harmony and vice versa. In this study, the level of group harmony is a measured as a score, ranging from 1 to 9. We measured the performance in terms of mean absolute error (MAE) for each activity and their weighted mean (overall). With MAE metric, the lower the value, the better the performance of the model. This is performed by extracting the generated text into numerical value, for example, \emph{\{``group harmony": ``4"\}} indicates that the predicted level of group harmony is 4. 

Cross-task is inherently challenging as it tests the boundaries of a model's generalization capabilities. This difficulty was further compounded in our evaluation by using the in-house dataset from a different domain, introducing a concurrent cross-domain challenge. Table \ref{tab:noc_harmony_results} shows the performance of group harmony analysis. We observed that all  model finetuned on OUC-CGE overfit to the task of group engagement analysis. Consequently, they are unable to generate text conformed to the format of group harmony analysis, thus, the result of ``Standard approach" model is not included in this table. To overcome this issue, we utilized the standard Qwen2.5-VL-3B-Instruct model for summarization step in our proposed models. This demonstrates the flexibility and robustness of \methodname compared to end-to-end modeling. As shown in Table \ref{tab:noc_harmony_results}, all our proposed model, except for ``Proposed - affective", demonstrates strong performance compared to Qwen2.5-VL-3B-Instruct baseline. Especially, ``Proposed - engagement"'s results are close or even surpassed those from the much larger Qwen2.5-VL-72B-Instruct. 

With the definition of group harmony above, we expected that the proposed model that uses individual affective states would perform well. However, this is not the case, as shown by the results in Table \ref{tab:noc_harmony_results}. Manual investigation of the generated individual affective states, valence, and arousal values indicated that they exhibited low accuracy. The observed inaccuracies are expected, considering two primary factors. First, large pre-trained models such as Qwen2.5-VL, even large models like Qwen2.5-VL-72B-Instruct, are known to struggle in affective computation tasks. Second, the individual analyzer model was finetuned on an out-of-domain dataset (OUC-CGE), which further compounded the difficulty of this task. These issues explained the low performance of ``Proposed - affective" model.

\section{CONCLUSIONS}
We proposed \methodname, a novel two-step approach for understanding social interaction dynamics, and tested the model on group engagement analysis. Our method is designed to first analyze individual-level cues and then synthesize them to form a prediction of group engagement level. To overcome the absence of individual cues ground-truth labels in existing dataset, we employed a knowledge distillation method. We conducted a comprehensive set of experiments to validate \methodname against several baseline and state-of-the-art models on the OUC-CGE dataset and our in-house dataset which will be made public. 

The experiment results demonstrate the effectiveness of our method. The models trained by our two-step approach achieved state-of-the-art results on OUC-CGE dataset with near-perfect performance. The results of ablation study indicate that non-verbal behavioral cues such as visual focus and gestures, as well as individual engagement, contribute the most to group engagement analysis. Furthermore, the proposed method also achieved strong cross-domain and cross-task generalization performance, reaching the performance of much larger models. This hightlights the flexibility and practical utility of \methodname for understanding group social dynamics. Finally, the proposed architecture is designed to leverage batch processing, which makes it practical for deployment in robotic applications.  

However, \methodname is limited by the input modality which relies only on video images, in contrast to the multi-modality of human-human social interaction. As the future work, we will extend the input modality of \methodname to include the audio and dialogue of social interaction to achieve better performance. To this end, we will first address the limitations of data scarcity in the field of group dynamics analysis. The group engagement analysis results on OUC-CGE indicate that the dataset may not be sufficiently complex or diverse. Additionally, the low performance of affective-based models show the limitations of VLMs in affective task. We therefore advocate for the development of a new dataset featuring fine-grained annotations for both individual cues such as emotions and group-level states such as group engagement and group harmony. Such a resource, particularly our in-house dataset which will be made public, would be invaluable for training more generalizable models and advancing the study of multiparty interactions. In terms of modeling, the proposed approach may encounter a computational bottleneck when applied to group interactions that involve a large number of participants. Therefore, more efficient analysis and summarization methods are required for such scenarios. 

\addtolength{\textheight}{-6cm}   








\bibliographystyle{IEEEtran}
\bibliography{my_references}

\end{document}